\documentclass[aps,prl,floatfix,twocolumn,showpacs]{revtex4}
\usepackage{epsfig}
\usepackage{verbatim}	 % for block comments \begin{comment} \end{comment}
\usepackage{amssymb}
\usepackage{bbold}
\usepackage{amsmath}
\usepackage{psfrag}
\usepackage{color}
\usepackage{ulem}
\newcommand{\eps}{\varepsilon}

\newcommand{\bi}{\bibitem}

\newcommand{\tD}{\tau_{\rm D}}

\newcommand{\NL}{N_{\rm L}}
\newcommand{\Nn}{N_{\rm n}}
\newcommand{\NR}{N_{\rm R}}
\newcommand{\NS}{N_{\rm S}}
\newcommand{\NSR}{N_{\rm SR}}
\newcommand{\NSL}{N_{\rm SL}}
\newcommand{\NC}{N_{\rm C}}

\begin{document}

% Patch REVTeX to prevent BibTeX from seeing endnotes as citations
% Insert just after REVTeX is loaded
%\makeatletter
%\let\@ORGREVTEXendnotemark\@endnotemark
%\let\@ORGREVTEX@makefnmark@cite\@makefnmark@cite
%\def\@endnotemark{\bgroup\@fileswfalse\@ORGREVTEXendnotemark\egroup}
%\def\@makefnmark@cite{\bgroup\@fileswfalse\@ORGREVTEX@makefnmark@cite\egroup}
%\makeatother

%\twocolumn[
\hsize\textwidth\columnwidth\hsize\csname@twocolumnfalse\endcsname

\title{Macroscopic Coherent Rectification in Andreev Interferometers}

\author{Jonathan Meair$^1$ and
Philippe Jacquod$^{1,2,3}$}
\affiliation{$^1$Physics Department, University of Arizona, Tucson, AZ 85721, USA \\$^2$College of Optical Sciences, University of Arizona, 
Tucson, AZ 85721, USA \\
$^3$D\'epartement de Physique Th\'eorique, Universit\'e de Gen\`eve,
CH-1211 Gen\`eve 4, Switzerland}

\vskip1.5truecm
\begin{abstract}
We investigate nonlinear transport through quantum coherent metallic conductors
contacted to superconducting components. We find that in certain geometries,
the presence of superconductivity generates a large, finite-average
rectification effect. Specializing to Andreev interferometers, we show that the direction and magnitude 
of rectification can be controlled by a magnetic flux tuning the superconducting phase 
difference at two contacts. In particular, this results in the breakdown of
an Onsager reciprocity relation at finite bias. The rectification current is macroscopic in that
it scales with the linear conductance, and we find that it
exceeds 5\% of the linear current at sub-gap biases of few tens of $\mu eV$'s.
\end{abstract}
% 74.45.+c Proximity effects; Andreev effect; SN and SNS junctions
% 73.23.-b Electronic transport in mesoscopic systems
% 74.78.Na Mesoscopic and nanoscale systems

\maketitle

The presence of superconductivity often magnifies quantum coherent
effects in transport. Examples include Aharonov-Bohm 
oscillations in the conductance~\cite{Pet93,Har96a,Naz96}
and in the thermopower~\cite{Eom98,Par03,Sev00,Vir04,Tit08,Jac10},
coherent backscattering~\cite{Bee95,Har98} and resonant tunneling~\cite{Goo08}.
The mechanism behind this 
enhancement can be traced back to Andreev reflection~\cite{And64}
which generates
new (diffuson-like) contributions to the transmission, that are
sensitive to different phases in the superconducting order parameter or to
external magnetic fluxes~\cite{Naz96,Sev00,Vir04,Tit08,Jac10,Eng11}. These contributions
are proportional to the number $N$ of transmission
channels. In purely metallic systems, quantum coherent 
effects are of order one or smaller, they are therefore
negligible in the limit $N \gg 1$ of large linear conductances~\cite{Imry}.

Novel quantum coherent effects in transport 
have recently been uncovered in the form of nonlinear contributions to 
the current-voltage characteristic. Of particular interest are 
contributions that are odd in a magnetic field $B$, 
$I_{\rm nl} = {\cal G}^{(2)}(B) V^2$ with ${\cal G}^{(2)}(B) = - {\cal G}^{(2)}(-B)$~\cite{San04,Spi04,And06,Zum06,Let06,Ang07}.
They originate from electronic interactions which, under finite applied biases,
modify the local potential landscape inside the conductor. 
The associated rectification coefficient ${\cal G}^{(2)} \propto \partial_E T_{ij} $ has been found to be 
proportional to the energy derivative of a transmission coefficient $T_{ij}$, and
in metallic quantum dots, it is accordingly 
sample-dependent, with a vanishing average and 
fluctuations decreasing with 
$N$, ${\rm var} (G^{(2)}) \propto N^{-2}$~\cite{San04,Spi04}.
Because $G^{(2)}$ is odd in $B$, its presence 
results in the breakdown of an Onsager
reciprocity relation~\cite{Ons31} at finite bias, $I(B,V) \ne I(-B,V)$.
According to Mott's law, at low temperature the thermopower is
also proportional to the energy derivative of the transmission~\cite{Ash67}. 
Therefore, the  question that naturally arises is
whether the enhancement of the thermopower observed in 
mesoscopic systems contacted to superconducting islands~\cite{Eom98,Par03,Sev00,Vir04,Tit08,Jac10},
translates into a similar magnification of nonlinear rectifying contributions to the
conductance. This is the problem we focus on in this manuscript. 

We investigate weakly nonlinear transport in coherent metals connected to 
superconducting contacts. 
We find that the presence of superconductivity renders 
the rectification current 
finite on average and macroscopically large -- in the sense that it scales with 
the linear conductance
$\langle {\cal G}^{(2)} \rangle = {\cal O}(N)$.
The emergence of a finite ${\cal G}^{(2)}$
does not require to break time-reversal symmetry in the metallic part of the system.
It takes place, for instance, 
for two superconducting contacts with phase difference $\phi \ne 0,\pi$, $\langle {\cal G}^{(2)} \rangle 
\propto \sin \phi$. The physics behind this 
effect is that, in Andreev systems, 
finite biases not only modify the local potential
landscape in the metal~\cite{Chri96,Pil02}, 
they also affect the electrochemical potential $\mu_{\rm sc}$ of the
superconductor. In the case of a superconducting island, 
$\mu_{\rm sc}$ adjusts itself to ensure current conservation, and therefore
transmission coefficients $T_{ij}(E,eU({\bf r}),\mu_{\rm sc})$ now depend on the absolute
energy $E$ of the charge carriers, the local potential landscape $U({\bf r})$ in the metal
and additionally on $\mu_{\rm sc}$.
Our key observation is that
rather generic hybrid systems can be devised where
Andreev reflection results in a large, finite-average derivative of $T_{ij}$ with respect to the
quasiparticle excitation energy $\varepsilon = E-\mu_{\rm sc}$,
$\langle \partial_\varepsilon T_{ij} \rangle = {\cal O}(N) \times  \sin \phi $. These 
contributions are similar to those giving a finite-average thermopower
in Andreev interferometers~\cite{Jac10,Eng11}.
The theory we are about to present
predicts maximal average rectification currents
amounting to 5--10\% of the linear current at still moderate, sub-gap biases
of 10--30 $\mu$V, and for which superconducting correlations persist over
distances of several microns. In purely metallic systems, fluctuating 
rectification effects on the order
of 2\% typically occur for biases in the range 0.1--1 mV~\cite{Zum06,Let06,Ang07}.

\begin{figure}[ht]
\psfrag{R} {$R$}
\psfrag{L} {$L$}
\psfrag{t}{$\delta \tau$}
\psfrag{D1}{$\Delta e^{-i \phi/2}$}
\psfrag{D2}{$\Delta e^{ i \phi/2}$}
%\psfrag{Gammas}{$\Gamma_{\rm n}, N_{\rm s}$}
\includegraphics[width=9cm]{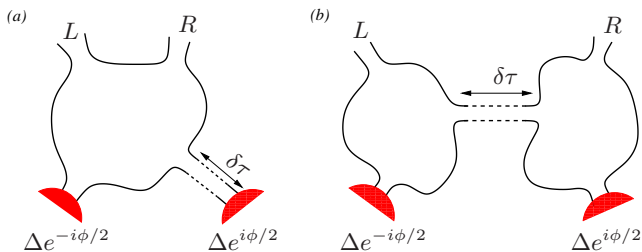}
\caption{\label{fig:models} (Color online) Sketch of the two Andreev interferometers we consider: 
(a) asymmetric single-cavity, and (b)
double-cavity interferometer. The red half circles represent the contacts to superconductors,
whose pair potentials are indicated.}
\end{figure}

We consider two models of Andreev interferometers,
where two metallic terminals indexed $i={\rm L,R}$
are connected to mesoscopic (either chaotic ballistic or 
disordered diffusive) quantum dots via leads
carrying $N_i\gg1$ transmission channels. The dots have no particular spatial symmetry and
are ideally connected to two $s$-wave superconducting contacts 
with pair potentials $\Delta e^{i\phi_i}$, each 
carrying ${\NS}_i$ channels. Physical properties depending
only on phase differences, we 
set $\phi_{\rm L}=-\phi/2$ and $\phi_{\rm R}=\phi/2$ with $\Delta \in {\cal R}$. 
We consider a single superconducting island with two contacts 
into which no current flows on time average in steady-state.
The models are sketched in Fig.~\ref{fig:models}. 
We consider the regime where the temperature is much smaller
than the pair potential, the latter being in its turn much smaller than the
Fermi energy, $T \ll \Delta \ll  E_{\rm F}$. At low bias, 
$eV \ll \Delta$, the quasiparticle excitation
energy is then always much smaller than $\Delta$. 
Accordingly, we assume perfect Andreev reflection at the superconducting contacts. 
 
Both our models are specifically devised to correlate the average time an electron takes on
its way from one lead to a superconducting contact, with the phase at that contact. 
This is achieved by the introduction of ballistic necks, which quasiparticles 
at the Fermi level cross in a time $\delta \tau$. These necks are indicated by dashed lines
in Fig.~\ref{fig:models}.
The way action and Andreev reflection
phases are correlated is easy to see by considering 
electrons at an excitation energy $\eps$, injected from the left 
terminal and Andreev reflected back to it. From Fig.~\ref{fig:models} we see that,
if Andreev reflection occurs at the right 
superconducting contact, these electrons acquire a total phase that is 
larger by an amount $2 \eps \delta \tau - \phi$ than if they hit 
the left superconducting contact.
Such correlations were shown in Ref.~\cite{Jac10} to generate large, finite-average
contribution to the thermopower for finite $\phi$. We show below that they also
result in a finite-average rectification.

The starting point of our analysis is the scattering theory formula for the 
electric current in terminal $i$~\cite{Cla96} (we set $\hbar, k_{\rm B} \equiv 1$
and express electric currents in units of $2e^2/h$)
\begin{align} \label{eq1}
I_i &= 
\int_0^\infty \frac{d\eps}{e} \;  \sum_j \sum_{\alpha, \beta} \alpha  \left( N_i \delta_{ij}\delta_{\alpha \beta} - T_{ij}^{\alpha \beta} \right) f^\beta_j (\eps) \, , \end{align}
with quasiparticle indices $\alpha,\beta=e, +1$ for electrons and
$h,-1$ for holes, the Fermi function for a $\beta$-quasiparticle
$f^\beta_j(\eps)=\left(\text{exp}\left\{\left[\eps - \beta e( V_j - V_{\rm sc})\right]/
T\right\}+1\right)^{-1}$ and 
the positive-defined quasiparticle excitation energy, $\eps=|E-eV_{\rm sc}|=|E-\mu_{\rm sc}|$. 
We introduced 
transmission coefficients $T^{\alpha \beta}_{ij}$
for a $\beta$-quasiparticle injected from lead $j$ to an $\alpha$-quasiparticle
exiting in lead $i$. In the presence of superconductivity,  transmissions will depend on 
(i) the energy $E$ of the injected quasiparticle, (ii) the local potential landscape 
$U({\bf r})$ on the quantum dot, and (iii) the electrochemical
potential $\mu_{\rm sc} = eV_{\rm sc}$ on the superconductor. We take
$\mu_{\rm sc}$ as our reference energy and express the transmission probabilities in terms of two energy differences, $T^{\alpha \beta}_{ij}(eU({\bf r})-\mu_{\rm sc},\eps)$, which describes how transmission is affected by the local potential landscape and the quasiparticle's excitation energy.

At low but finite bias we expand Eq.~(\ref{eq1}) to quadratic order in 
$V_j-V_{\rm sc}$ and write the current as
\begin{eqnarray}
I_i = \sum_j G_{i j}^{(1)} (V_j-V_{sc})+\sum_{j,k} G^{(2)}_{i j k} (V_j-V_{\rm sc})(V_k-V_{\rm sc}) \, . \label{current} \end{eqnarray}
The linear, $G^{(1)}_{ij}$, and quadratic, $G^{(2)}_{ijk}$ conductances are
given by
\begin{subequations}
\label{nonlinear_cond}
\begin{eqnarray} 
\label{nonlinear_conda}
G^{(1)}_{ij} &=& \int_{0}^{\infty} d\eps \; \left(-\partial_\eps f \right) \; g_{ij} \, ,\\
\label{nonlinear_condb}
G_{ijk}^{(2)} &= &\frac{1}{2}\int_{0}^{\infty} d\eps \; \left(-\partial_\eps f \right) \; \left(e \partial_\eps b_{ij}\delta_{jk} + 2\partial_{V_k} g_{ij}\right) \, , \,\,\,\,\,\,
\end{eqnarray}
\end{subequations}
with
\begin{subequations}
\begin{eqnarray} 
g_{ij} &=&\sum_{\alpha, \beta} 
\alpha \beta \left( N_i \delta_{ij}\delta_{\alpha \beta} - T_{ij}^{\alpha \beta} \right) \, ,\\
b_{ij} &=& \sum_{\alpha, \beta} 
\alpha \left( N_i \delta_{ij}\delta_{\alpha \beta} - T_{ij}^{\alpha \beta} \right) \, .
\end{eqnarray}
\end{subequations}
The linear conductance has been calculated earlier
(see e.g. Refs.~\cite{Jac10,Eng11}), and we focus our attention on the
nonlinear term. 
The second term in the parenthesis in 
Eq.~(\ref{nonlinear_condb}) gives the screening contribution to the nonlinear conductance, self-consistently generated by the voltage biases and the Coulomb interaction~\cite{Chri96}. It can be rewritten 
as~\cite{Chri96,San04}
\begin{align}
\left(\frac{\partial g_{ij}}{\partial V_k}\right) = \int d{\bf r} 
\frac{\delta g_{ij}}{\delta \left(U({\bf r})-V_{\rm sc}\right)} 
\left(\frac{\partial \left(U({\bf r})-V_{\rm sc}\right)}{\partial V_k}\right) \, .
\end{align}
For metallic mesoscopic 
cavities the derivative of the transmission with respect to the local potential is random from one ensemble member to another, furthermore, for large $N$, it is not correlated to the local potential 
fluctuations~\cite{San04}. This is not modified by the presence of superconductivity, 
therefore, $\left<\partial_{V_k} g_{ij}\right>=0$.
The screening term has no average effect on the current, and we neglect it 
from now on.
The first term in Eq.~(\ref{nonlinear_condb}) is the bare contribution to the nonlinear conductance~\cite{Chri96}. In contrast to the purely metallic case, it depends on the derivative of the transmission
with respect to the quasiparticle excitation energy $\eps$
counted from the chemical potential of the superconductor. Below we find that 
the bare term gives a dominant, finite-average contribution to the nonlinear conductance.
We thus rewrite Eq.~(\ref{current}) as
\begin{eqnarray}
I_i = \sum_j G_{i j}^{(1)} (V_j-V_{\rm sc})+\sum_{j} G^{(2)}_{i j} (V_j-V_{\rm sc})^2 \, , \label{current2} 
\end{eqnarray}
with 
\begin{eqnarray}
\label{nonlinear_condb2}
G_{ij}^{(2)} & = &\frac{e}{2}\int_{0}^{\infty} d\eps \; \left(-\partial_\eps f \right) \; \partial_\eps b_{ij} \, .
\end{eqnarray}

\begin{figure}[ht]
\includegraphics[width=6.5cm]{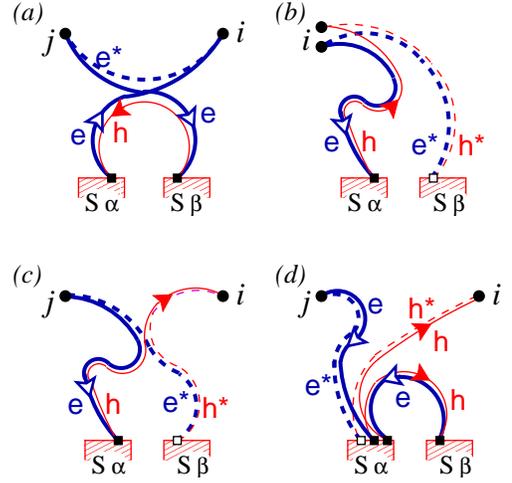}
\caption{\label{fig:processes}(Color online) Semiclassical diagrams
that give the dominant phase-sensitive 
contributions to the rectification coefficient ${\cal G}^{(2)}$, to leading order in $\NS/\Nn$~\cite{Jac10}.
(a) Contribution to 
$\langle T_{ij}^{\rm ee}\rangle$, and (b,c,d) contributions to
$\langle T_{ij}^{\rm he} \rangle$.
Blue (red) lines indicate electrons (holes) trajectories, while
dashed lines indicate complex-conjugated amplitudes.
Normal leads are labelled $i,j$ while superconductors are labelled $S\alpha,S\beta$. Contributions to $\langle T_{ij}^{\rm hh}\rangle$ [$\langle T_{ij}^{\rm eh} \rangle$] are obtained from (a)
[(b,c,d)] by substituting $e \leftrightarrow h$ everywhere.
}
\end{figure}

Eq.~(\ref{current2}) expresses electric currents through the normal leads as a function of the
superconducting chemical potential $\mu_{\rm sc}=e V_{\rm sc}$. In steady-state, the latter is self-consistently
determined by current conservation $I_{\rm L}=-I_{\rm R}$ at the normal leads.
The next step is thus to determine $\mu_{\rm sc}$ and insert its value into
Eq.~(\ref{current2}). One gets an explicitly gauge invariant expression
\begin{align}\label{eq:current_gauge}
I_L &= {\cal G}^{(1)} \; V +
{\cal G}^{(2)} \;V^2 \, ,
\end{align}
with the bias voltage  $V=V_{\rm L} - V_{\rm R}$ and
\begin{subequations}\label{eq:g1g2}
\begin{eqnarray}
{\cal G}^{(1)} & = & (G^{(1)}_{\rm LL} G^{(1)}_{\rm RR} -G^{(1)}_{\rm LR} G^{(1)}_{\rm RL} )\big/\sum_{kl}G^{(1)}_{kl} \, , \\
{\cal G}^{(2)} & = & \sum_l (
G^{(2)}_{LR} G^{(1)}_{{\rm R} l}
-
G^{(2)}_{RR} G^{(1)}_{{\rm L} l}) 
\nonumber \\
&& \times \sum_k (G^{(1)}_{k\rm L} - G^{(1)}_{k\rm R} ) \big/
 \big(\sum_{kl} G^{(1)}_{kl}\big)^2 \, .
\end{eqnarray}
\end{subequations}
It is easily checked that the expression for ${\cal G}^{(1)}$ reproduces the result of Ref.~\cite{Cla96}.
To calculate the rectification coefficient ${\cal G}^{(2)}$
we follow the trajectory-based semiclassical approach of Refs.~\cite{Jac10,Whi09} and compute
the dominant contributions to leading order in the ratio $N_{\rm SL,SR}/N_{\rm L,R}$ of the
total number of transport channels in the left and right 
superconducting contacts ($N_{\rm SL}$ and $N_{\rm SR}$) and in the normal
contacts ($N_{\rm L}$ and $N_{\rm R}$). The corresponding diagrams are shown in Fig.~\ref{fig:processes}.

Eqs.~(\ref{eq:current_gauge}--\ref{eq:g1g2}) apply to both interferometers shown in Fig.~\ref{fig:models},
however,
the coefficients $G_{ij}^{(1,2)}$ are geometry-dependent and we now calculate them in both cases,
starting with the 
asymmetric single-cavity interferometer.
For the interferometer of Fig.~\ref{fig:models}a, we find, to leading order in $N_{\rm L,R}$, 
$N_{\rm SR,SL}$ and $N_{\rm SL,SR}/N_{\rm L,R}$, 
\begin{subequations}
\begin{eqnarray}\label{eq:lincond}
\left<{\cal G}^{(1)}\right> &=& \frac{\NL \NR}{\NL + \NR} \, ,
\\
%%%%
\label{eq:asym_scav}
\left<{\cal G}^{(2)} \right> &=&
2e   \sin \phi \,  \frac{\NL \NR \NSL \NSR (\NR-\NL)}{(\NL + \NR)^4}   \,  I_a \, , \qquad
\end{eqnarray}
\end{subequations}
with a thermal damping integral
\begin{equation}\label{eq:ia}
I_a = \int_0^\infty {\rm d} \eps \; (-\partial_\eps f) \; \partial_\eps \left[ \frac{\sin(2 \eps \delta \tau)}{1 + (2 \eps \tD)^2} \right].
\end{equation}
The value of the integral $I_a$ decreases with the temperature, $T$, and the dwell time, $\tD$,
through the cavity. When $ T  \ll  \tD^{-1}$ one has $I_a=2\pi T \, \delta \tau^2 \, \text{csch}(2\pi T 
\delta \tau)$. For a given temperature, it is largest when $
\delta \tau \simeq 0.3 /T$ [when $\pi T \delta \tau$ coth$(2\pi T \delta \tau)=1$]. We see from Eq.~(\ref{eq:asym_scav}) that,
when $\delta \tau$ is nonzero, a finite average rectification current flows. This current 
is odd in $\phi$. Both finite $\phi$ and $\delta \tau$ are required for this current to occur, 
because they both are necessary to correlate action 
and Andreev phases. This correlation is key to obtaining a finite-average $\partial_\eps b_{ij}$
in Eq.~(\ref{nonlinear_condb2}). Eq.~(\ref{eq:asym_scav}) further shows that when $N_i = {\cal O}(N) \gg 1$, 
$i$=L, R, SL, SR, and for sufficiently asymmetric normal terminals, $|N_{\rm L}-N_{\rm R}| \gg 1$,
the rectification current is macroscopically large, ${\cal G}^{(2)} = {\cal O}(N)$. 

We next present our results 
for the double-cavity interferometer. We find, again to leading order,
\begin{subequations}
\begin{eqnarray}
\left<{\cal G}^{(1)}\right> &=&  \NC + \frac{2\NSL \NSR}{\NSL+\NSR} \, ,
\\
%%%%
\label{eq11b}
\left<{\cal G}^{(2)}\right> &=&
2e \sin \phi \frac{(\NSL-\NSR)^2 \NSL \NSR \NC}{(\NSL+\NSR)^2 \NL \NR}  \, I_b \, , \qquad
\end{eqnarray}
\end{subequations} 
where $\NC \ll N_{\rm L,R}$ is the number of channels in the neck connecting the two cavities
and the thermal damping integral is given this time by
\begin{equation}
I_b = \int_0^\infty d\eps \; (-\partial_\eps f) \; \partial_\eps \left[ \text{Im}\left\{\frac{\text{exp}(2 i \eps \delta \tau)}{
(1 + 2 i \eps \tD)^2} \right\}\right] \, .
\end{equation}
Here,
for simplicity, we took the same dwell time, $\tD$, for both cavities. 
>From Eq.(\ref{eq11b}), we see
that  a macroscopic rectification effect also occurs in this geometry -- under similar conditions as above,
i.e. that $N_i = {\cal O}(N) \gg 1$ for $i$=L, R, C, SL and SR, and sufficiently 
asymmetric superconducting contacts, $|\NSL-\NSR| \gg 1$ -- 
and that when $T  \ll \tD^{-1}$, rectification effects in the two
geometries of Fig.~\ref{fig:models} have the same thermal damping, $I_b=I_a$. 

\begin{figure}[ht]
\includegraphics[width=8.5cm]{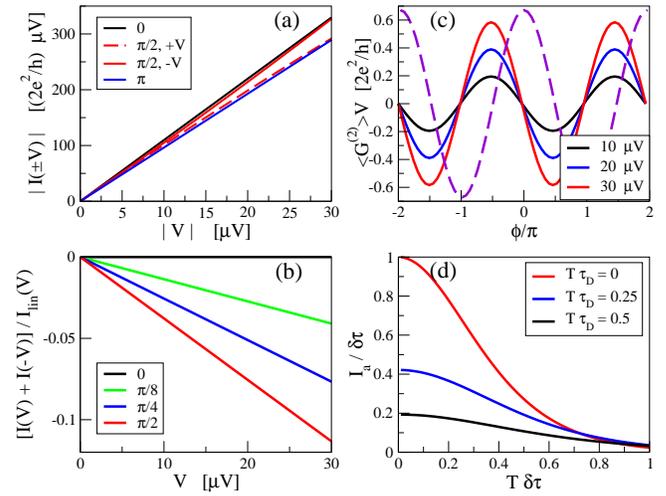}
\caption{\label{fig:data} (Color online) Rectification in an asymmetric single-cavity interferometer with 
$\NL=110$, $\NR=10$, $\NSL=\NSR=30$, $T=100$mK, $\delta \tau=0.3 /T$, $T \ll \tD^{-1}$.
(a) Electric current at $\pm V$ vs. the bias voltage $|V|$. At $\phi=0$ and $\pi$, $I(V)=-I(-V)$
and there is no rectification. At $\phi=\pi/2$, rectification is maximal (red curves). 
A subdominant oscillating contribution to the linear current (see Eq.(3a) in Ref.~\cite{Jac10}),
not included in Eq.(\ref{eq:lincond}), is taken into account here.
(b) Relative current asymmetry $I(V)+I(-V)$ normalized with the linear current
$I_{\rm lin}(V) = G^{(1)} V$
for four different superconducting phase differences. 
(c) Sample-averaged rectification conductance $\langle {\cal G}^{(2)} \rangle V$ 
vs. $\phi$ for three different bias voltages. The oscillating part of the linear conductance 
is shown for comparison (dashed line).
(d) Thermal
integral $I_a$ of Eq.~(\ref{eq:ia}) giving the damping of $\langle {\cal G}^{(2)} \rangle$ with
temperature for the single-cavity model of Fig.~\ref{fig:models}a.}
\end{figure}

We illustrate our results in Fig.~\ref{fig:data} for the  asymmetric
single-cavity model. We first show
the current as a function of applied bias in Fig.~\ref{fig:data}a, for $\phi=0$ and $\phi=\pi/2$.
We see that a rectification effect of more than 5\% occurs at $\phi=\pi/2$ and bias voltage of
30 $\mu$V. This is rendered more evident in Fig.~\ref{fig:data}b, which 
shows the relative current asymmetry $[I(V)+I(-V)]/I_{\rm lin}(V)$, normalized by the 
linear current $I_{\rm lin} = {\cal G}^{(1)} V$ as a function of bias voltage. We see that
at still moderate biases (well below the superconducting gap of Al, and corresponding to a
coherence length $v_{\rm F}/eV$ ranging from tens to hundreds of $\mu$m for GaAs 2DEG to 3D metals),
 the rectification effect
exceeds 5\%. 
We next show in Fig.~\ref{fig:data}c the rectification current as a function of
$\phi$ for three different voltages $V=10$, 20, and 30$\mu$V. In contrast to 
the mesoscopic rectification effects in metallic quantum dots which are random 
in an applied magnetic field~\cite{San04,Spi04,Zum06},
we see that the presence of superconductivity induces a regular behavior of $\langle {\cal G}^{(2)} \rangle$
as a function of $\phi$, with the magnitude of the effect increasing with bias. Finally, the damping of the rectification with temperature is 
illustrated in Fig.~\ref{fig:data}d. \\
 
Our approach to weakly nonlinear transport is closely related to the one pioneered by B\"uttiker and Christen \cite{Chri96}. One important difference is that we here took advantage of the presence of superconductivity
to Taylor-expand the currents in voltages measured from the superconducting potential $V_{\rm sc}$.
This directly enforces gauge invariance at our level of approximation, where the screening
term in Eq.~(\ref{nonlinear_condb}) is neglected. 
Current conservation is furthermore satisfied in our treatment by 
unitarity of the scattering matrix, $\sum_{i,\alpha} T_{ij}^{\alpha \beta}=N_j$, and by the 
condition (self-consistently determining $V_{\rm sc}$) that  no current enters the superconducting 
island on time average in steady state. 
In Ref.~\cite{Chri96}, voltages are taken from an arbitrary potential as there is no superconductor. 
In that case gauge invariance is only satisfied after a self-consistent 
determination of the local potential landscape $U({\bf r})$ and of the dependence of transmission
coefficients on external voltage biases via the latter.

We have presented a theory for weakly nonlinear transport in hybrid metallic/superconducting systems and shown that there can be a finite average ${\cal O}(N)$ rectification for such systems. We found that,
in contrast to purely metallic mesoscopic systems, 
the presence of superconductivity generates potentially large, ${\cal O}(N)$, finite-average 
rectification effects. The latter can furthermore be tuned in magnitude and direction by an external
magnetic flux.
Alternatively, we note that this effect  leads
to the breakdown of an Onsager relation, with (still in units of $2e^2/h$)
$I(\phi,V)-I(-\phi,V)= 4e  \, \sin \phi [\NL \NR \NSL \NSR (\NR-\NL)/(\NL+\NR)^4 ] I_a V^2$ for the asymmetric single-cavity model and 
$I(\phi,V)-I(-\phi,V)= 4 e  \, \sin \phi [(\NSL- \NSR)^2 \NSL \NSR \NC/ (\NSL+\NSR)^2 \NL \NR] I_b V^2$
for the double-cavity model.
We expect the rectification effect we predict to be experimentally testable in 
Andreev interferometers such as those of Refs.~\cite{Har96a,Eom98,Par03}.

This work was supported by the NSF under grants
DMR-0706319 and PHY-1001017
and by the Swiss Center of Excellence MANEP. It is our pleasure to
thank M. B\"uttiker and P. Stano for discussions.


\begin{thebibliography}{99}

\bi{Pet93} V.T. Petrashov, V.N. Antonov, P. Delsing, and T. Claeson, Phys. Rev. Lett. {\bf 70},
347 (1993).

\bi{Har96a} S.G. den Hartog, C.M.A. Kapteyn, B.J. van Wees, T.M. Klapwijk, and G. Borghs,
Phys. Rev. Lett. {\bf 77}, 4954 (1996).

\bi{Naz96} Y.V. Nazarov and T.H. Stoof, Phys. Rev. Lett. {\bf 76}, 823 (1996).

\bi{Eom98} J. Eom, C.-J. Chien, and V. Chandrasekhar, 
Phys. Rev. Lett. {\bf 81}, 437 (1998).

\bi{Par03} A. Parsons, I.A. Sosnin, and V.T. Petrashov, Phys. Rev. B {\bf 67}, 140502(R) (2003).

\bi{Sev00} R. Seviour and A.~F. Volkov, Phys. Rev. B {\bf 62}, R6116 (2000).

\bi{Vir04} P. Virtanen and T. Heikkil\"a, 
Appl. Phys. A {\bf 89}, 625 (2007).

\bi{Tit08} M. Titov, Phys. Rev. B {\bf 78}, 224521 (2008).
  
\bibitem{Jac10} Ph. Jacquod, R. S. Whitney, Europhys. Lett. {\bf 91} 67009 (2010).

\bibitem{Bee95} C.W.J. Beenakker, J.A. Melsen, and P.W. Brouwer,
Phys. Rev. B {\bf 51}, 13883 (1995).

\bibitem{Har98} S.G. den Hartog, B.J. van Wees, Yu.V. Nazarov, T.M. Klapwijk, and G. Borghs,
Physica B {\bf 249}, 467 (1998).

\bibitem{Goo08} M.C. Goorden, Ph. Jacquod, and J. Weiss,
Phys. Rev. Lett. {\bf 100}, 067001 (2008); 
Nanotechnology {\bf 19}, 135401 (2008).

\bi{And64} A.F. Andreev, Sov. Phys. JETP 19, 1228 (1964).

\bibitem{Eng11} T. Engl, J. Kuipers, and K. Richter,
Phys. Rev. B {\bf 83}, 205414 (2011).

\bibitem{Imry} Y. Imry, {\it Introduction to Mesoscopic Physics}, 3$^{\rm rd}$ Ed.
(Oxford University, Oxford, 2008).

\bibitem{San04} D. Sanchez and M. B\"uttiker, Phys. Rev. Lett. {\bf 93}, 106802 (2004).

\bibitem{Spi04} B. Spivak and A. Zyuzin, Phys. Rev. Lett. {\bf 93}, 226801 (2004).

\bibitem{And06} A.V. Andreev and l.I. Glazman, Phys. Rev. Lett. {\bf 97}, 266806 (2006).

\bi{Zum06} D. Zumb\"uhl, C.M. Marcus, M.P. Hanson, A.C. Gossard,
Phys. Rev. Lett. {\bf 96}, 206802 (2006).

\bi{Let06} R. Leturcq, D. Sanchez, G. G\"otz, T. Ihn, K. Ensslin, D.C. Driscoll, 
and A. C. Gossard,  Phys. Rev. Lett. {\bf 96}, 126801 (2006).

\bi{Ang07} L. Angers, E. Zakka-Bajjani, R. Deblock, S. Gu\'eron, H.
Bouchiat, A. Cavanna, U. Gennser, and M. Polianski,
Phys. Rev. B {\bf 75}, 115309 (2007).

\bi{Ons31} L. Onsager, Phys. Rev. {\bf 38}, 2265 (1931).

\bi{Ash67} N.W. Ashcroft and N.D. Mermin, {\it Solid-State Physics} 
(Saunders College Publishing, Philadelphia, 1967).

\bibitem{Chri96} T. Christen and M. B\"uttiker, Europhys. Lett. {\bf 35}, 523 (1996).

\bi{Pil02} S. Pilgram, H. Schomerus, A.M. Martin, and M. B\"uttiker,
Phys. Rev B {\bf 65}, 045321 (2002).

\bibitem{Cla96} N. R. Claughton and C. J. Lambert, Phys. Rev. B {\bf 53}, 6605 (1996).

\bibitem{Whi09} R. S. Whitney, P. Jacquod, Phys. Rev. Lett. {\bf 103}, 247002 (2009).




\end{thebibliography}
\end{document}